\documentclass[a4paper,12pt,fleqn]{article}
\usepackage{amsfonts}
\usepackage{amsmath}
\usepackage{amsopn}
\usepackage{amsthm}
\usepackage{amssymb}
\usepackage{color}
\usepackage{graphicx}
\usepackage{rotating}
\usepackage[english]{babel}
\usepackage[utf8]{inputenc}
\usepackage{xspace}
\usepackage{float}
\usepackage{setspace}
\usepackage{dsfont}
\usepackage{booktabs}
\usepackage[T1]{fontenc}
\usepackage{anysize}
\usepackage[labelfont=bf,labelsep=period]{caption}
\usepackage{subfig}
\usepackage{tabularx}
\usepackage{times}

\allowdisplaybreaks

\begin{document}
\title{Robust Bayesian Modeling with Adaptive Posterior FDR Control for Large-Scale Data}

\author{Yoshiko Hayashi$^{}$}

\maketitle

\thispagestyle{empty}

\begin{center}
\noindent $^{}$Osaka Metropolitan University
\end{center}

\begin{abstract}
Controlling the false discovery rate (FDR) is a critical challenge in large-scale data analysis, particularly in the presence of outliers. A common practice involves imposing a Student-$t$ distribution to eliminate the influence of outliers. Here, we developed a robust Bayesian analysis based on heavy-tailed modeling, applied it to large-scale studies in Bayesian inference, and performed diagnoses for detecting outliers using the posterior predictive $p$-value ($ppp$). In addition, we propose an adaptive method to decide the level of the posterior false discovery rate. We demonstrated the utility of our methods using gene expression data for colorectal cancer. We suggest an adaptive method to determine it using an estimated ratio of true null genes using Storey's $q$-value method.
\end{abstract}


\section{Introduction}

Differential gene expression experiments in typical two-color DNA microarrays have been extensively examined in recent decades. 
Outliers in microarray analysis can be managed by modeling the error term with a heavy-tailed distribution.
For instance, \cite{R4} solved the outlier problem with a robust Bayesian hierarchical model using Student's $t$-distribution for the gene expression data. 
Other studies have also suggested using heavy-tailed distributions for managing outliers in gene expression data analysis. For example, \cite{R15} noted that gene expression data exhibit a heavier distribution than a normal distribution, and \cite{R9} and \cite{R7} also adopted heavy-tailed distributions.
Furthermore, \cite{R14} managed with outliers using empirical Bayesian correction for variances. We applied the robust Bayesian modeling on a large-scale dataset.

During the analysis for false discovery rate (FDR), as \cite{R18} noted, misspecification of the cut point for the calculation of the posterior probability of the difference in gene expression between the two groups leads to failed control of the FDR. To address this issue, we adopted an adaptive criterion to calculate the posterior FDR. 

The manuscript is organized as follows. In Section 2, our model setting and two methods of detecting outliers are presented. Section 3 outlines an adaptive criterion for controlling the posterior FDR. In Section 4, simulation studies for examining the performance of the robust modeling and on the comparison of the behavior of the proposed posterior FDR and the real FDR are outlined. In Section 5, we illustrate the utility of our model and method using gene expression data for colorectal cancer. Section 6 presents a discussion of the results of the study.

\section{Model}

Bayesian heavy-tailed modeling and conflict resolution have considerably progressed over the past few decades. In this study, we applied the Student-$t$ linear regression model, a heavy-tailed modeling approach examined previously by \cite{R3} and \cite{R77}. We applied the Bayesian Student-$t$ linear regression model to analyze gene expression data. 

 For the linear regression analysis, we assumed that the error terms are independent and followed a $t$-distribution with degrees of freedom $d$, mean 0, and the scale parameter $\delta$, $u_{gi} \sim  t_{(d)}(0,\delta).$ The parameter $\gamma_{s}$ is a state variable $\gamma_{s}=\{0, 1\}$.
 We assumed $G$ genes with $S$ subjects (g=1, \ldots , G, s=1,\ldots, S). For gene expression data, $g$th gene, and $s$th subject, $y_{gs}$, the model is outlined below:
\begin{eqnarray}
y_{gs}&= &\beta_{0,g} + \beta_{1,g}  \gamma_{s} +u_{gs}. 
\end{eqnarray}
As priors, we adopted the independent Jeffreys prior derived by Fonseca et al. (2008). In the model, we assumed that the priors are independent and specified the degrees of freedom. 
\begin{equation}
\left\{
\begin{array}{@{\,}ll}
\beta_{g}  \stackrel{D}{\sim} Uniform,\\
\sigma  \stackrel{D}{\sim} 1/\sigma. \\
\end{array}
\right.
\end{equation}
where, $\beta_{g}=\{\beta_{0,g}, \beta_{1,g} \}$.

With our criteria, \cite{R77} showed that the Student-$t$ modeling can partially reject outliers, where the number of outliers is $L$, when the condition $L \cdot d<(S-L-2)$ is satisfied.

\subsection{Outlier Detection}
\label{sec:3}
The posterior predictive distribution is used for model diagnosis. When a model explains the data, the posterior probability will be high. For example, \cite{R6} assessed the goodness of fit using the posterior predictive distribution and showed the conflict between outliers and non-outliers using a Gaussian model. In addition, \cite{R5} derived conflict measures for the hierarchical model.\cite{R11} also presented a conflict measure based on the posterior distribution. Our model uses the posterior predictive distribution from the opposite viewpoint. 

We focused on the model behavior where outliers cannot be adequately explained by the heavy-tailed model. If the observation is not adequately explained by the posterior predictive distribution, we infer that the outlier is eliminated by the robust modeling. We propose two statistics based on the posterior predictive distribution: an overall diagnostic to assess whether a dataset includes any outliers, and an individual diagnostic to assess whether a given observation is an outlier. 

The posterior predictive $p$-value ($ppp$) was calculated from $A$ iterations of a Markov chain Monte Carlo (MCMC) process as follows:
\begin{eqnarray}
ppp&=&P(D(y^{\mathrm{rep}};\theta)\leq D(y^{\mathrm{obs}};\theta) \mid y) \\ \nonumber
&\doteq &\frac{1}{A}\sum^A_{i=1} I \{D(y^{\mathrm{rep}};\theta)\leq D(y^{\mathrm{obs}};\theta) \mid y \}
\end{eqnarray}
where $D(y|\theta)$ denotes a discrepancy measure based on the posterior predictive distribution, and $y^{\mathrm{obs}}$ and $y^{\mathrm{rep}}$ represent the observed data and replication, respectively.

\subsubsection{Individual outlier detection}
\label{sec:5}
In addition, we propose the following individual diagnostic to detect each outlier:

\begin{eqnarray}
ppp(y)=P(y^{(\mathrm{rep})} \leq y^{(\mathrm{obs})} \mid data).
\end{eqnarray}

According to \cite{R10} and \cite{R11}, the predictive distribution of $y$ is obtained by cross-validation without the data concerned to avoid conservative results. Thus, without outliers, we expect that the distribution is uniform. For the purpose of detecting outliers, we use the statistics as follows.
This statistic would be close to zero for a positive outlier and close to one for a negative outlier.

\subsubsection{Overall outlier detection}
\label{sec:4}
A small $ppp$ value indicates that the model does not explain the corresponding observation. In this case, the statistics computed using data including the outliers differ from the posterior statistic. We defined the posterior predictive $p$-value to detect whether a dataset includes any outliers, as follows:

\begin{eqnarray}
ppp(S^2)=P(S^{2(\mathrm{rep})} \leq S^{2(\mathrm{obs})} \mid data)
\end{eqnarray}
where $S^{2(\mathrm{rep})}$ and $S^{2(\mathrm{obs})}$ denote the variances of the replication and observed data, respectively.

\cite{R8} assumed the common scale parameter for the statistics to avoid conservative results. Here, we calculated it using each dataset, which indeed led to conservative results, as our purpose was to detect outliers. Under this approach, the statistics of a dataset containing the outliers would be close to zero.\\

\section{Posterior FDR}
\cite{R13} proposed the FDR based on the posterior distribution, which is further explained by \cite{R12}. \cite{R17} also investigated a spatial model that uses the Bayesian posterior FDR. In Bayesian inference, a non-zero positive critical value must be specified to calculate the posterior probability of differential gene expression. Here, we propose an adaptive method to determine the critical value. 

Consider a multiple testing problem in Bayesian analysis with $G$ tests ($H_{0g}$ $(g=1, \ldots ,G)$), where $G_{0}$ and $G_{1}$ denote the numbers of unknown true nulls and non-nulls, respectively. Our decisions yield $(G-R^*)$ nulls (non-discovery cases) and $R^*$ non-nulls (discovery cases) as shown in Table 1.

The multiple testing problems depend on our decision ($d_i$) and the unknown truth ($t_i$). When the $i$-th test corresponds to a true non-null hypothesis, $t_i$ takes 1: $t_i=1$. When the test corresponds to a non-null hypothesis, $t_i =0$. The decision is based on the $s_i$ statistic. Therefore, $d_i=1$ if $s_i \geq s$, and $d_i=0$ if $s_i < s$.

 According to \cite{R12}, in Bayesian modeling, the truth is replaced by the posterior probability or expected true non-null and null probabilities.
Let $p_{Bg}=P(t_{g}=1 \mid data)$ denote the marginal posterior probability that gene $g$ is differentially expressed.
Therefore, $p_{Bg}$ associated with ($d_g=0$) represents the posterior probability of a false negative, and $1-p_{Bg}$ associated with ($d_g=1$) represents the posterior probability of a false discovery. The decision ($d_g$) is based on $p_{Bg}$ using a predetermined threshold, $p_{cut}$ (see \cite{R20}. Therefore, $d_g=1$ if $p_{Bg} \geq p_{cut}$, and $d_g=0$ if $p_{Bg} < p_{cut}$.

The posterior probabilities of the true non-null and null possibilities in each test are listed in Table 1. The decision is based on the predetermined criterion $p_{cut}$ as \cite{R13} suggested that the level of the posterior FDR needs to be determined first, and the level of $p_{cut} $ needs to be calculated.

\begin{table}[h]
\begin{center}
\caption{Multiple tests in Bayesian modeling}
\begin{tabularx}{130mm}{lllc}
\toprule
& Non-discovery & Discovery & Total\\
& $(p_{Bg}<p_{cut})$ & $(p_{Bg} \geq p_{cut})$ & \\
\midrule
Null&
$\sum_{g \in G(p_{Bg} < p_{cut})} (1-p_{Bg}) $ & $\sum_{g \in G(p_{Bg} \geq p_{cut})}  (1-p_{Bg})$ & $G_0$\\
Non-null & $\sum_{g \in G(p_{Bg} < p_{cut})}  p_{Bg} $ & $\sum_{g \in G(p_{Bg} \geq p_{cut})}  p_{Bg}$ & $G-G_0$\\
\midrule
& $G-R^*$ & $R^* $ & $G$\\
\bottomrule
\end{tabularx}
\end{center}
\end{table}

Among the $R$ cases, when $V$ are incorrectly categorized as non-null, ordinal FDR is defined as $E(V/R)$ for positive $R$. According to \cite{R17}, the posterior FDR can be calculated as follows:
\begin{eqnarray}
FDR_{post}=\frac{1}{R^*} \sum_{g \in G(p_{Bg} \geq p_{cut})} (1-p_{Bg}),
\end{eqnarray}
where $R^*$ denotes the number of discoveries that are group members satisfying $(p_{Bg} \geq p_{cut})$. From a Bayesian perspective, the decisions are made when the posterior probability is exceeds 0.5. Thus, $p_{cut}>0.5$ is required.

In terms of the parameter concerned ($\theta$), $p_{Bg}$ is determined as follows: 
\begin{eqnarray}
p_{Bg}^{} \equiv Pr\{|\theta_g|>c \mid data \} \hspace{1cm} \textrm{for} \hspace{0.5cm} c > 0.
\end{eqnarray}
Thus, $p_{Bg}$ crucially depends on the critical value ($c$). 

\subsection{Adaptive critical value}
\label{sec:6}
Here, we outline the following method to specify the level of the critical value ($c$).
The best approach involves matching the number of true null genes ($G_0$) with the number of posterior null genes ($\hat{G_0}=\sum_{g \in G(p_{Bg} \geq p_{cut})}  (1-p_{Bg})$). Hence, the critical value should be selected under the condition described as follows:
\begin{eqnarray}
G_0=\sum_{g \in G(p_{Bg} \geq p_{cut})}  (1-p_{Bg}).
\end{eqnarray}
However, as the number of true null genes $G_0$ remains unknown, we propose an adaptive method for estimating $G_0$ and determining a critical value.

According to \cite{R16}, it is reasonable to set the critical value as zero, where the mean for null genes is located at 0. Thus, for the first step, to create the posterior distribution and obtain $p_{Bg0}$ for the null, we adopted a critical value of 0, with a one-sided probability as follows:
\begin{eqnarray}
p_{Bg0} \equiv Pr\{\theta>0 \mid data \}.
\end{eqnarray}
Since the distribution is based on the null, when the non-null has a positive effect, $p_{Bg0}$ is at or near zero, and when the non-null has a negative effect, $p_{Bg0}$ is at or near unity. Thus, we estimated the null ratio after adjusting $p_{Bg0}$ to a two-sided test method, converting $p_{Bg0}$ into $Tp_{Bg0} =1-0.5\cdot|p_{Bg0}-0.5|$. We then applied Storey's $q$-value method, which is provided by \cite{R20}, to the distribution of $Tp_{Bg0}$.

\section{Simulation Results}
\label{sec:9}
In this section, we demonstrate three aspects. First, to detect outliers using the posterior predictive $p$-value ($ppp$), we examined the behavior of $ppp$ in the presence of outliers. Second, we present the simulated results for the estimated null ratio for the case with and without outliers, and then we present the results of the calculated adaptive critical value based on the estimated null ratio. Third, we compared the posterior FDR for the case with and without outliers.

In the estimation model, we adopted a $t$ distribution with three degrees of freedom as the likelihood: $u_{gi} \sim t_{(3)}(0,\delta)$, and used the prior distributions of parameters in the model $\beta_{i,g}$, which were assumed to be uniform. The prior distribution of the scale parameter $\delta$ was assumed to be the independent Jeffreys prior given as $p(\delta)= 1/\delta$ (see \cite{R2}), as shown in the model section.

\subsection{Simulation Data}
\label{sec:M}

We used the error term with a $t$-distribution: $u_{gi} \sim t_{(3)} (0, 1)$. The parameters for generating samples were set as $\gamma_{i}=0$ for state 1 and $\gamma_{i}=1$ for state 2.
Outliers are generated by adding 100 to $y_{gi}$. We set two outliers for each state. Thus, each dataset included four outliers. The number of datasets with outliers (among the 1,000 datasets) is listed in Table 2.

\begin{table}[h]
\begin{center}
\caption{Numbers of datasets with outliers out of 1,000 datasets}
\begin{tabularx}{110mm}{lrr}
\toprule
& \quad\quad\quad\quad\quad\quad   Null & \quad\quad\quad\quad\quad\quad Non-null\\
\midrule
$\pi_0=0.90$ & 100 & 10\\ 
$\pi_0=0.95$ & 100 & 5\\ 
$\pi_0=0.99$ & 100 & 1\\ 
\bottomrule
\end{tabularx}
\end{center}
\end{table}

\subsection{Outliers Detection}
\label{sec:10}

Figure 1 shows the histograms of the posterior predictive probability to detect outliers. Panels (a) and (c) of Figure 1 show histograms of the posterior predictive probability ($ppp(y_{gi})$) for the simulated data null ratio, $\pi_0=0.90$. 

As shown in Figure 1 (a), $ppp(y)$ values followed a uniform distribution for the dataset excluding outliers. This suggests that the model performed well with the simulated data and did not find any significant outliers. However, Figure 1 (c) reveals the presence of outliers because the mode is clearly separated from the uniform distribution and exhibits a value approaching approximately zero, which indicates some positive outliers in our simulation dataset.

Figure 1 (b) and (d) show the results of $ppp(S^2)$. Both histograms reveal a clearly conservative distribution, since we did not use a common parameter to save calculation time. The $ppp(S^2)$ values for the model with outliers, shown in Figure 1(d), exhibit a clear mode around 0 as expected. By observing the $ppp$ of the scale parameter, we detected the number of datasets containing outliers as the mode around 0. Therefore, we computed $ppp(y)$ in the first step to determine the number of outliers and $ppp(S^2)$ in the second step as a whole. The subsequent step can be used to diagnose the number of datasets that include outliers.

\begin{figure}[ht]
\begin{center}
\subfloat[$ppp(y)$ ]{
\includegraphics[width=0.35\textwidth]{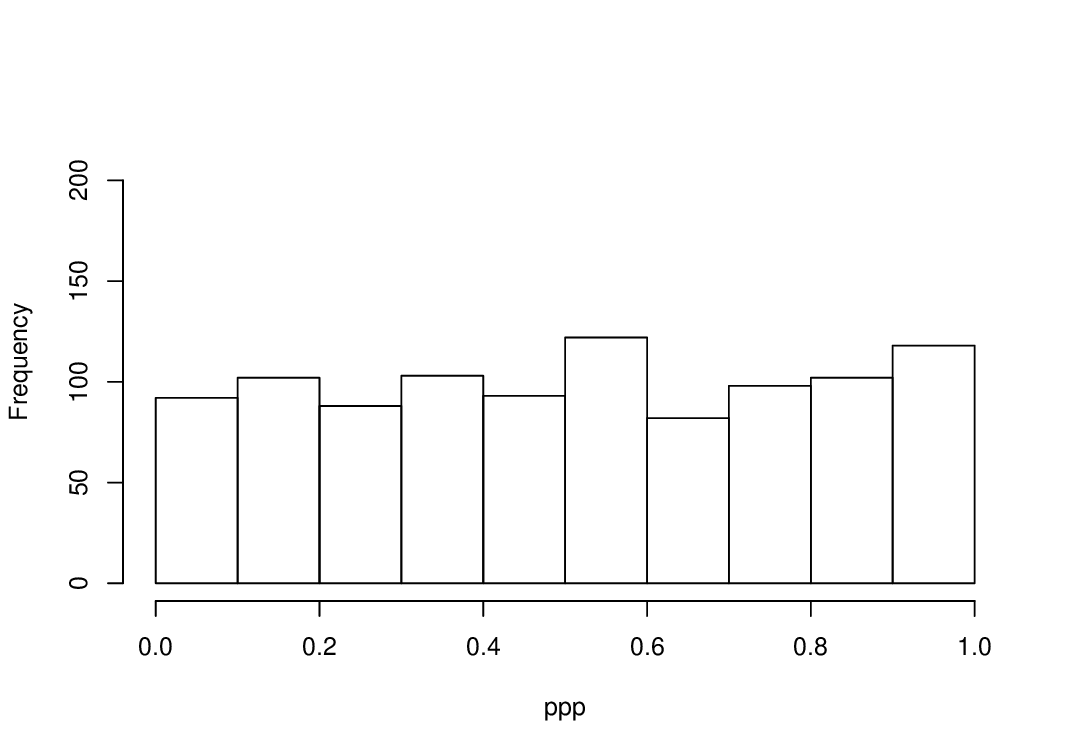}}
\subfloat[$ppp(S^2)$ ]{
\includegraphics[width=0.35\textwidth]{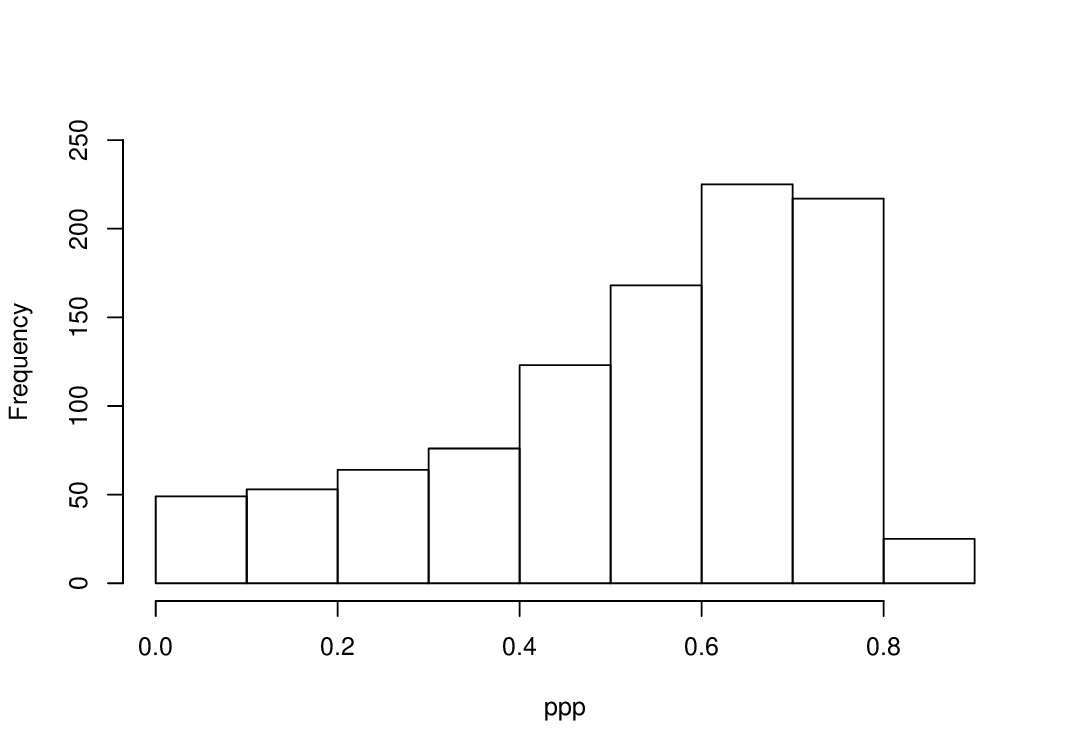}} \\
\subfloat[$ppp(y)$ with outliers ]{
\includegraphics[width=0.35\textwidth]{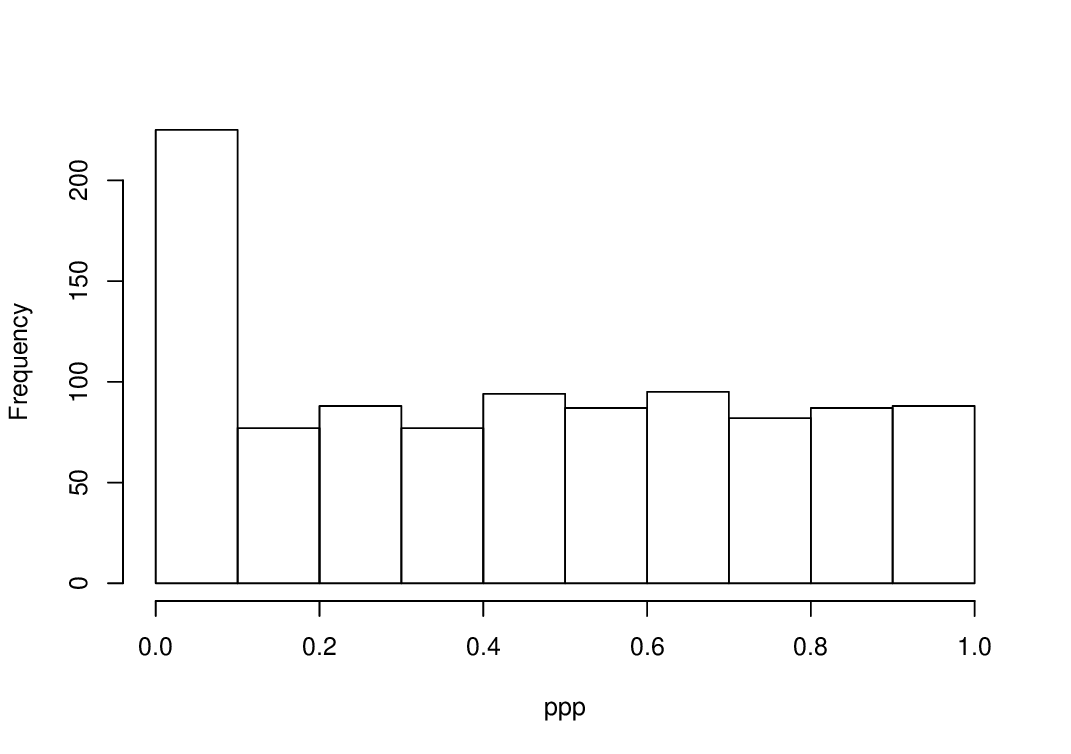}}
\subfloat[$ppp(S^2)$ with outliers ] {
\includegraphics[width=0.35\textwidth]{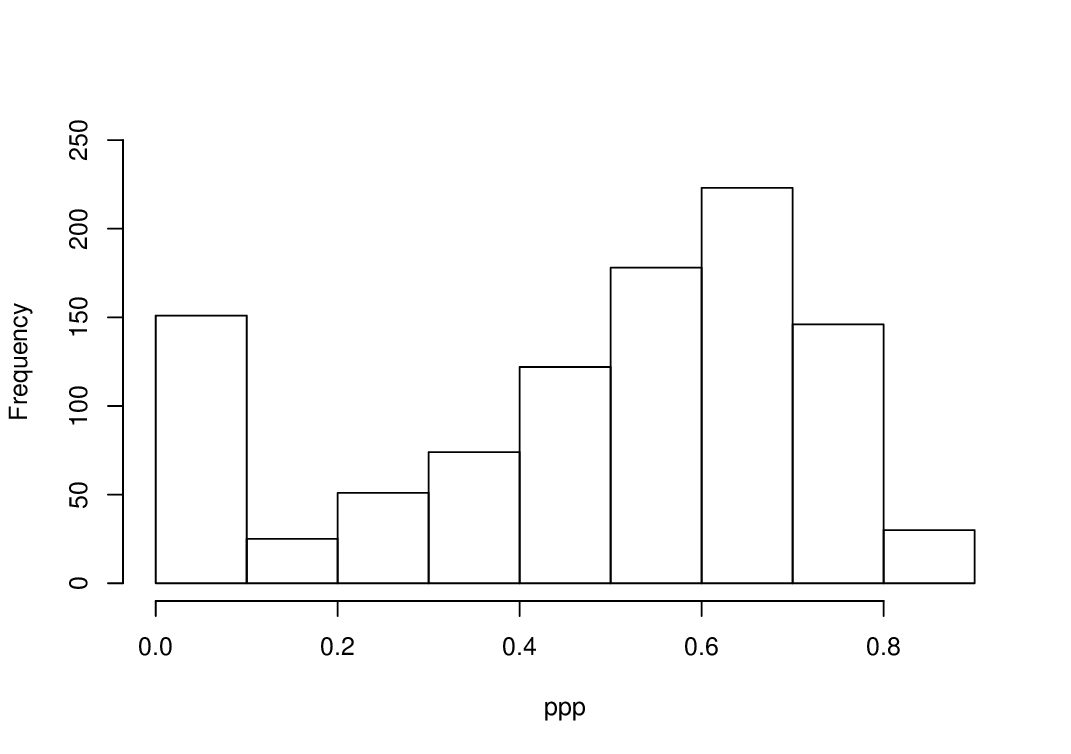}}
\caption{Posterior predictive $p$-values $ppp(y_i)$ with the cross-validation method and $ppp(S^2)$ for the simulated data null ratio, $\pi_0=0.90$. Panels (a) and (b) show the results of the model without outliers, and panels (c) and (d) show the results of the model with positive outliers.} \vspace{0cm}
\end{center}
\label{fig:1}
\end{figure}

\subsection{Adaptive critical value}
\label{sec:11}
To obtain the estimated null ratio, $\hat{G_0}$. We conduct the following procedure. We obtained $p_{Bg0}$ from 1,000 samples. After transforming the mode to a probability with two-sided tests, we applied Storey's $q$-value method. The results are listed in Table 3. The natural spline is fitted in the 0--0.5 range in increments of 0.05. The results show that the estimates are slightly larger than the true null ratio, and especially the dataset with outliers has larger estimates than the true ratio. This indicates that the estimate induces conservative FDR, especially for the dataset with outliers.\\

\begin{figure}[ht]
\begin{center}
\begin{minipage}[ ]{0.20\linewidth}
\end{minipage}\\
\subfloat[$\pi_0$ = 0.90  ]{
\includegraphics[width=0.35\textwidth]{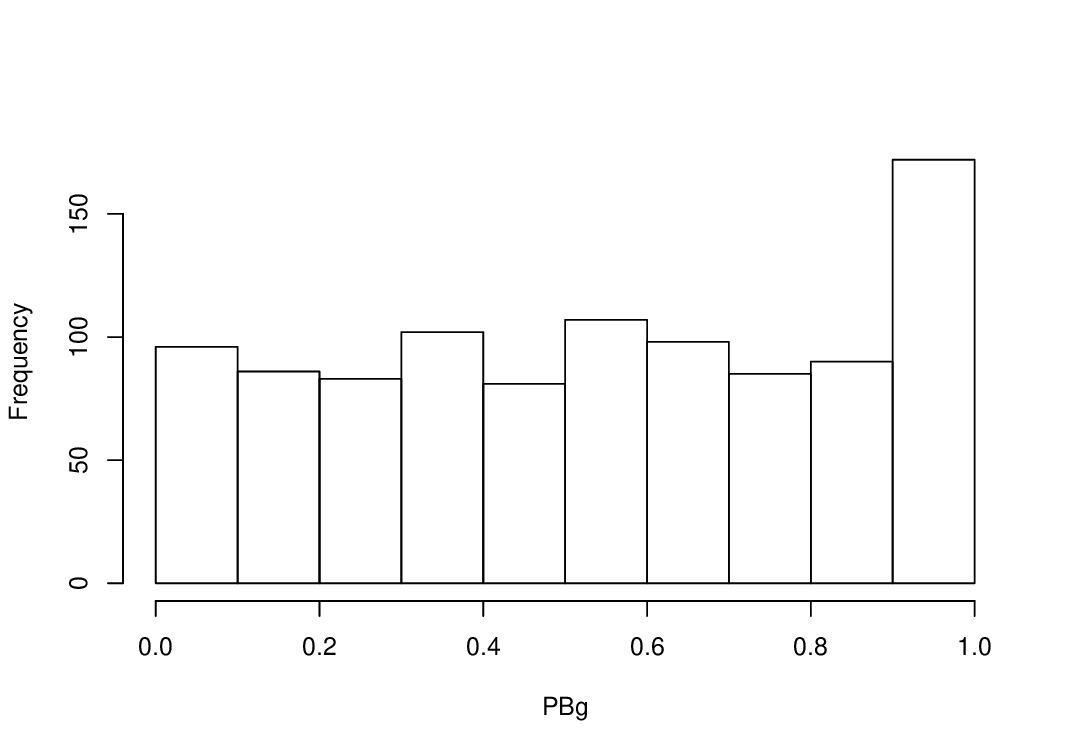}}
\subfloat[$\pi_0$ = 0.95 ]{
\includegraphics[width=0.35\textwidth]{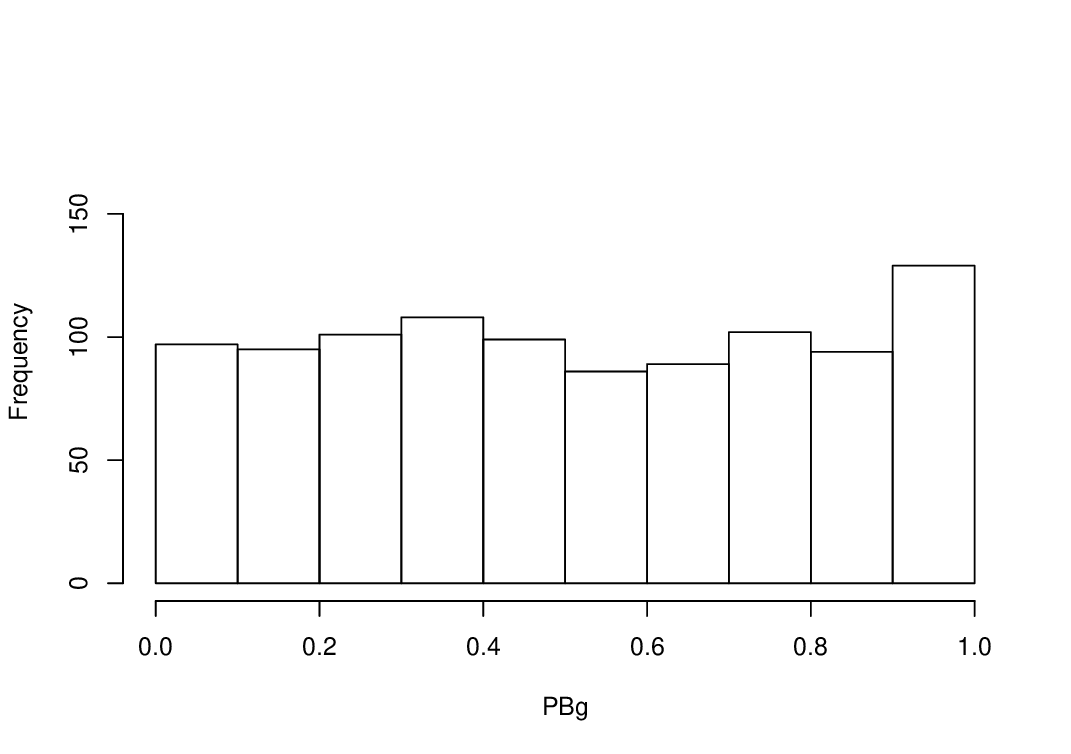}} 
\subfloat[$\pi_0$ = 0.99  ]{
\includegraphics[width=0.35\textwidth]{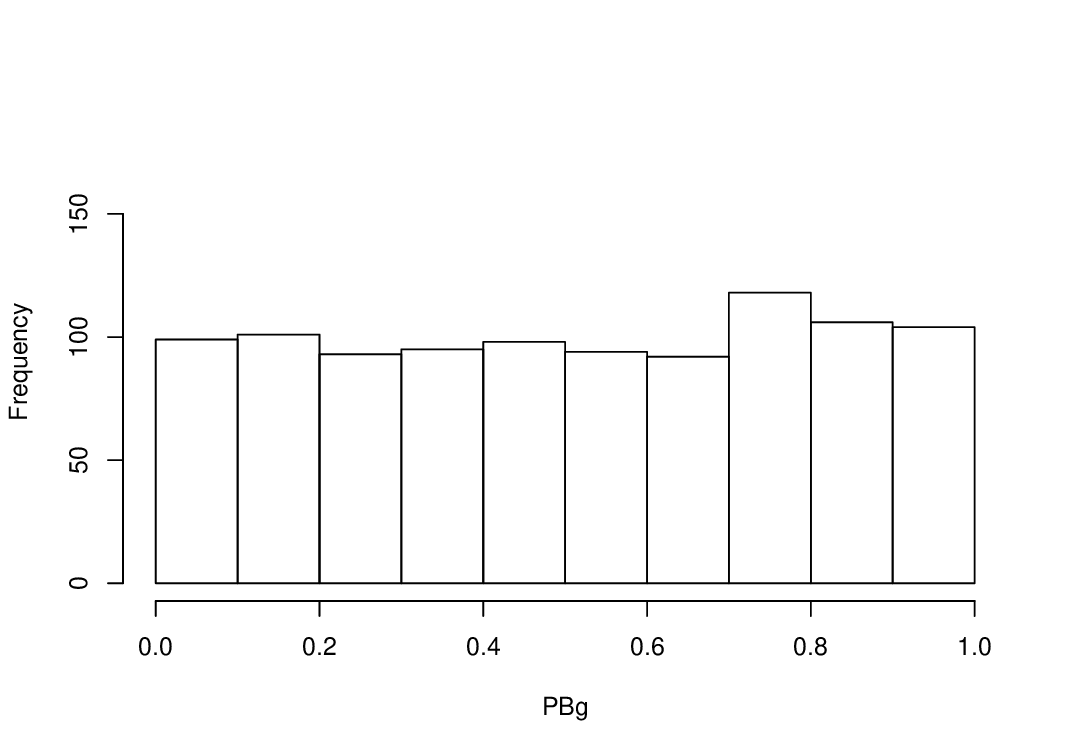}}\\
\vspace{1cm}
\begin{minipage}[ ]{0.27\linewidth}
\end{minipage}\\
\subfloat[ $\pi_0$ = 0.90 with outliers ] {
\includegraphics[width=0.35\textwidth]{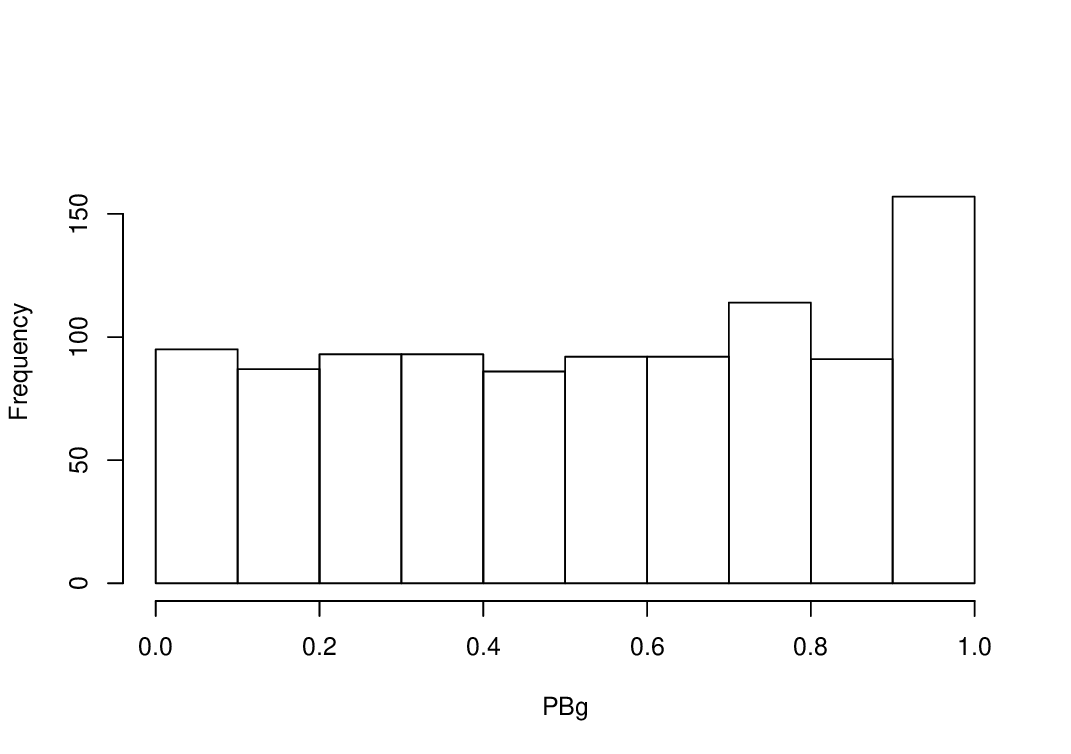}}
\subfloat[ $\pi_0$ = 0.95  with outliers] {
\includegraphics[width=0.35\textwidth]{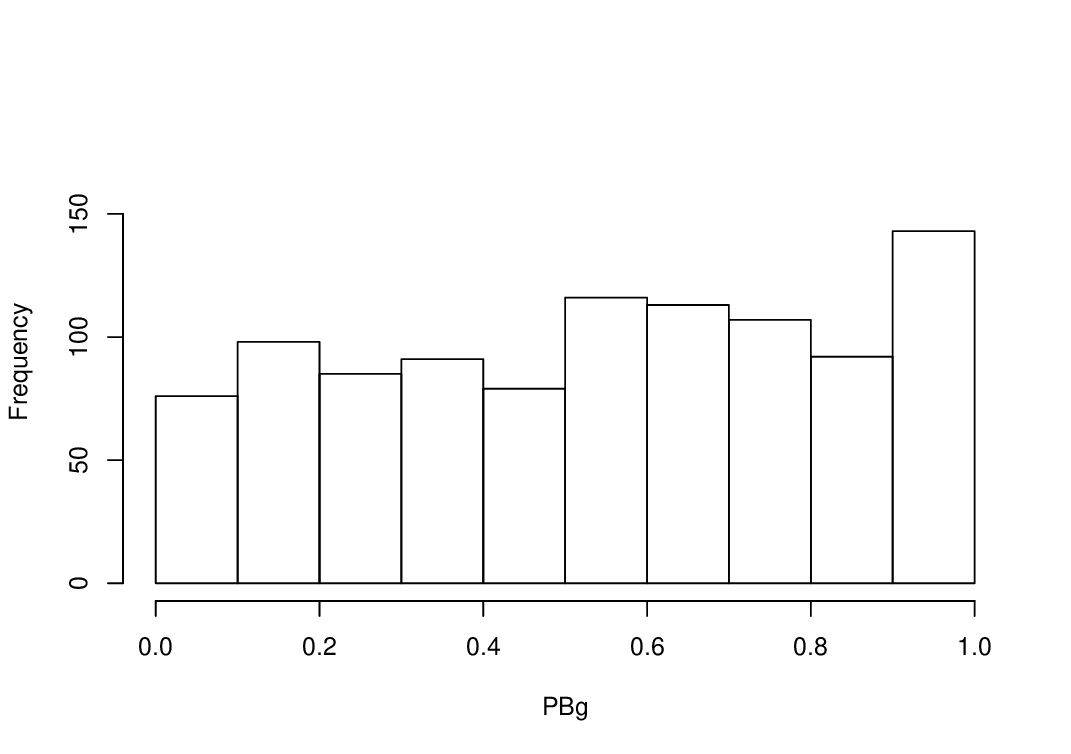}}
\subfloat[ $\pi_0$ = 0.99 with outliers  ] {
\includegraphics[width=0.35\textwidth]{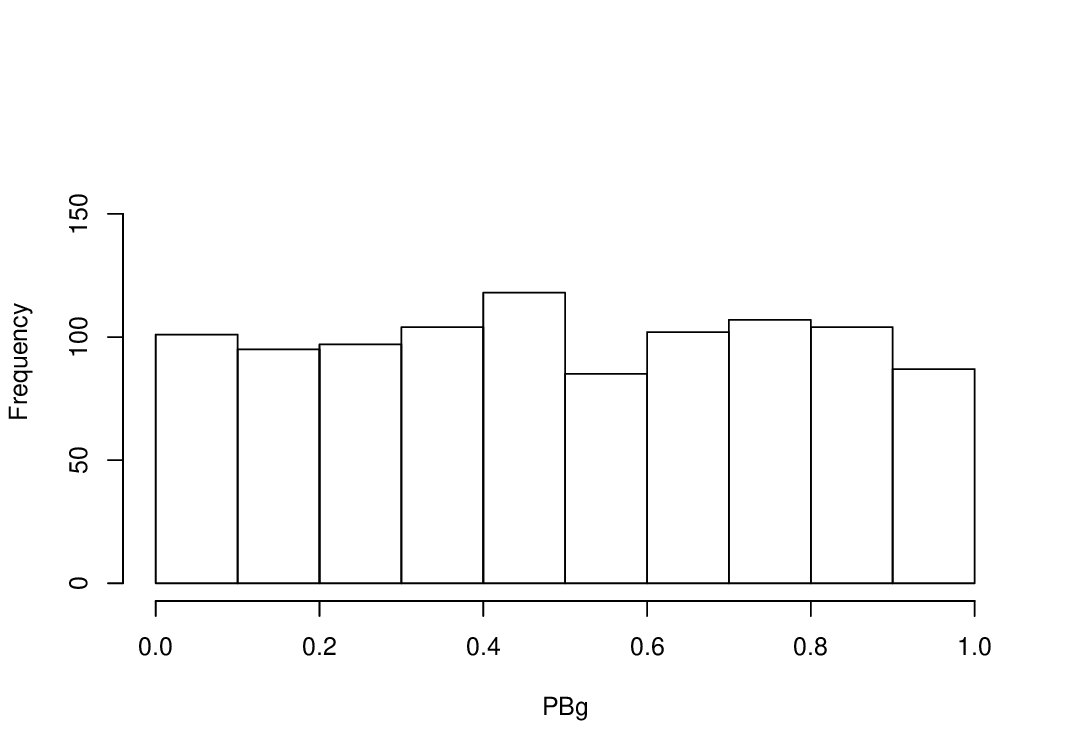}}\\
\caption{Histogram of $P_{Bg0}$. Panels (a), (b), and (c) show the results of the model without outliers, and panels (d), (e), and (f) show the results of the model with positive outliers.} \vspace{0cm}
\end{center}
\label{fig:2}
\end{figure}

\begin{table}[h]
\begin{center}
\caption{Estimated null ratio $\hat{G_0}$}
\begin{tabularx}{110mm}{lrr}
\toprule
&\quad\quad\quad Without outliers &\quad\quad\quad With outliers\\
\midrule
$\pi_0=0.90$ & 0.928 & 0.933\\ 
$\pi_0=0.95$ & 0.963 & 0.993\\ 
$\pi_0=0.99$ & 0.988 & 1.000\\ 
\bottomrule
\end{tabularx}
\end{center}
\end{table}

To obtain the adaptive critical value ($c^*$) used in equation (7), we set the critical value to meet $\sum_{g \in G(p_{Bg} \geq p_{cut})} (1-p_{Bg}) $ to $\hat{G_0}$, which is obtained in the above section.. First, we calculated the posterior number with critical values from 0.1 to 2.0 in increments of 0.1. Subsequently, we fitted a cubic smoothing spline on them and determined the critical value. The results are presented in Table 3. We used the critical value to obtain the posterior FDR.

\begin{table}[h]
\begin{center}
\caption{Adaptive critical value ($c^*$)}
\begin{tabularx}{110mm}{lrr}
\toprule
&\quad\quad\quad Without outliers &\quad\quad\quad With outliers\\
\midrule
$\pi_0=0.90$ & 1.21 & 1.24\\ 
$\pi_0=0.95$ & 1.37 & 1.46\\ 
$\pi_0=0.99$ & 1.81 & 1.98\\ 
\bottomrule
\end{tabularx}
\end{center}
\end{table}

\subsection{Posterior FDR}
\label{sec:P}
Figure 3 shows log-linear plots of the counted and posterior FDRs as functions of the number of discovery datasets for various thresholds of $p_{cut}$. For these simulations, we set the true null ratio to calculate $P_{Bg}$ here.

As highlighted by \cite{R16}, when the critical values are set equal to a true (simulated) null ratio, the posterior FDR behaves relatively similarly to the real FDR and is slightly conservative, especially at low discovery numbers. Notably, the FDRs of models with outliers are lower than those of models with no outliers because the heavy-tailed model leads to a larger variance in the posterior distribution. \\

\rm
\begin{figure}[ht]
\begin{center}
\begin{minipage}[ ]{0.20\linewidth}
\end{minipage}\\

\subfloat[$\pi_0$ = 0.90  ]{
\includegraphics[width=0.35\textwidth]{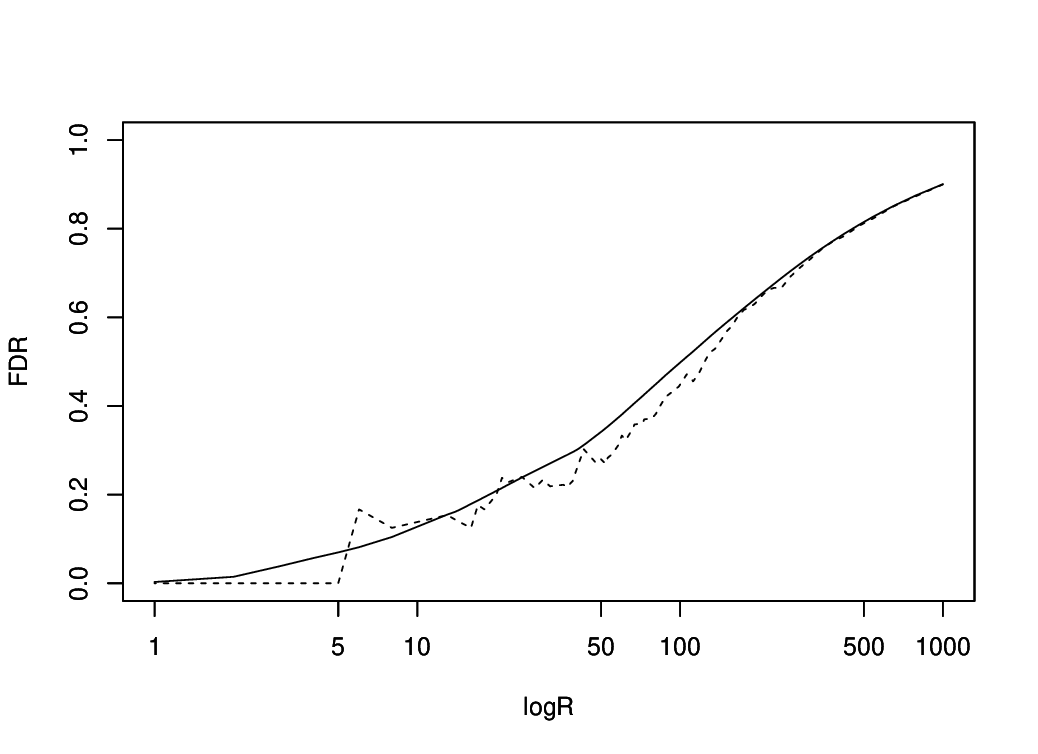}}
\subfloat[$\pi_0$ = 0.95 ]{
\includegraphics[width=0.35\textwidth]{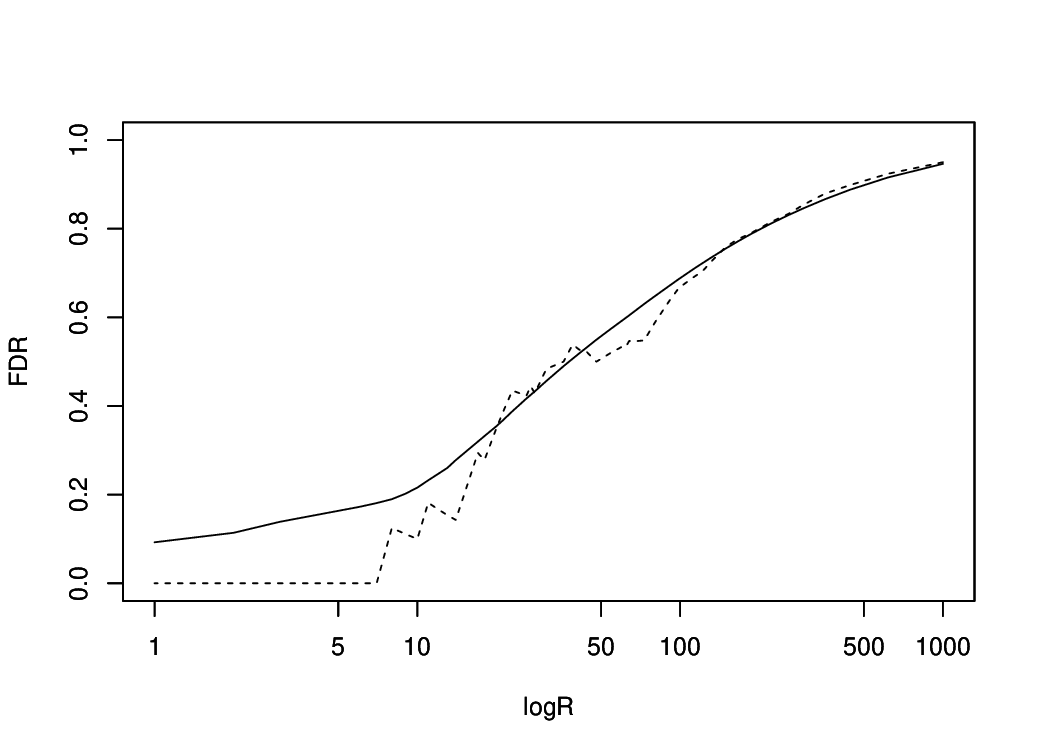}} 
\subfloat[$\pi_0$ = 0.99 ]{
\includegraphics[width=0.35\textwidth]{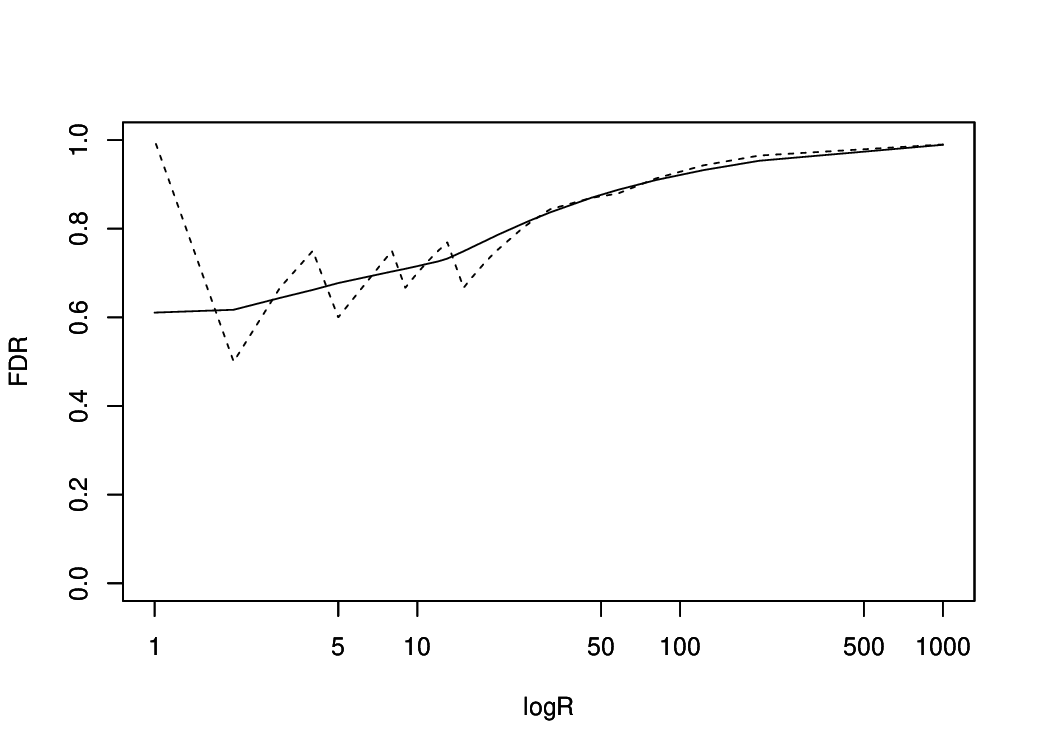}}\\
\vspace{1cm}
\begin{minipage}[ ]{0.17\linewidth}
\end{minipage}\\
\subfloat[ $\pi_0$ = 0.90 with outliers ] {
\includegraphics[width=0.35\textwidth]{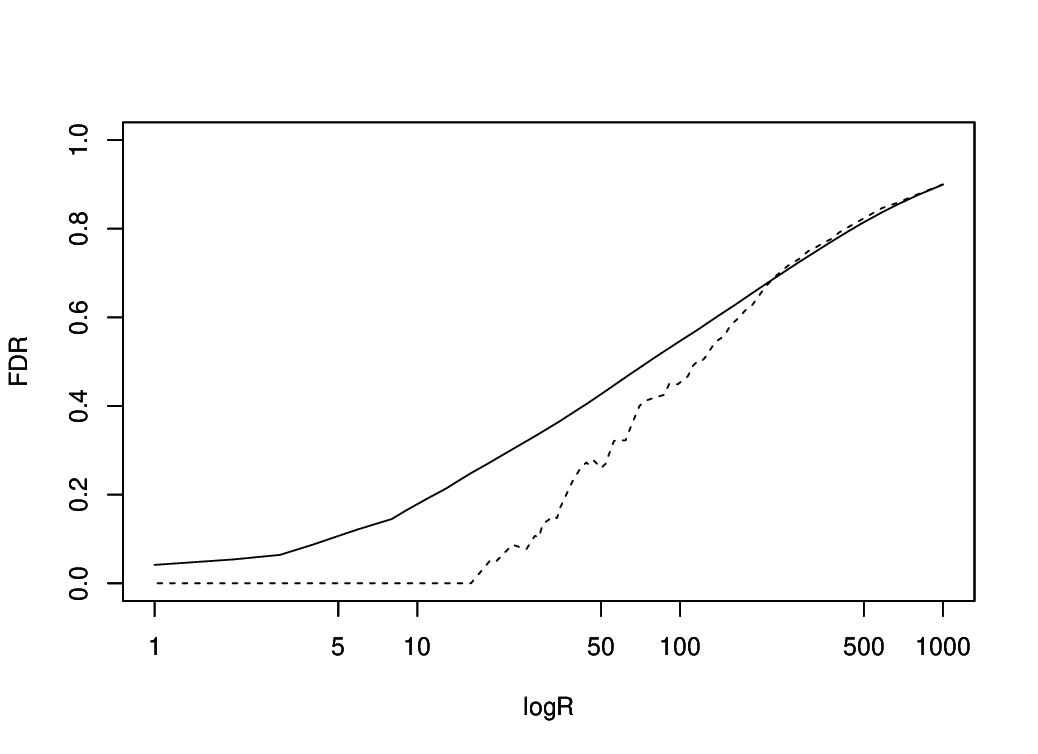}}
\subfloat[ $\pi_0$ = 0.95 with outliers ] {
\includegraphics[width=0.35\textwidth]{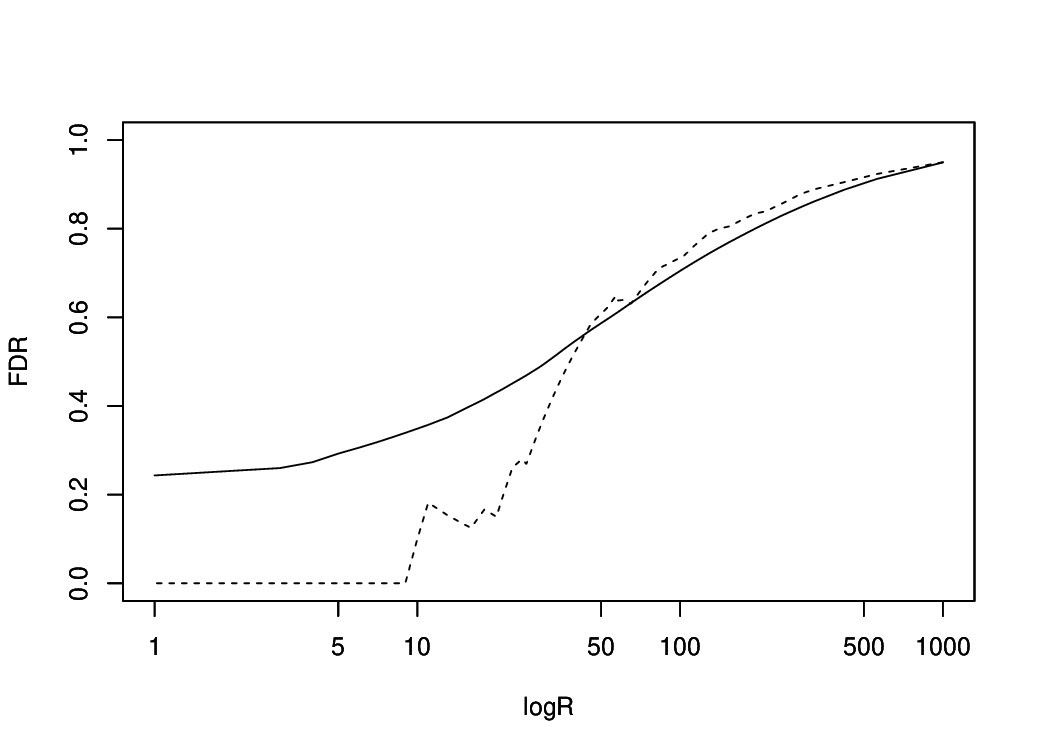}}
\subfloat[ $\pi_0$ = 0.99 with outliers  ] {
\includegraphics[width=0.35\textwidth]{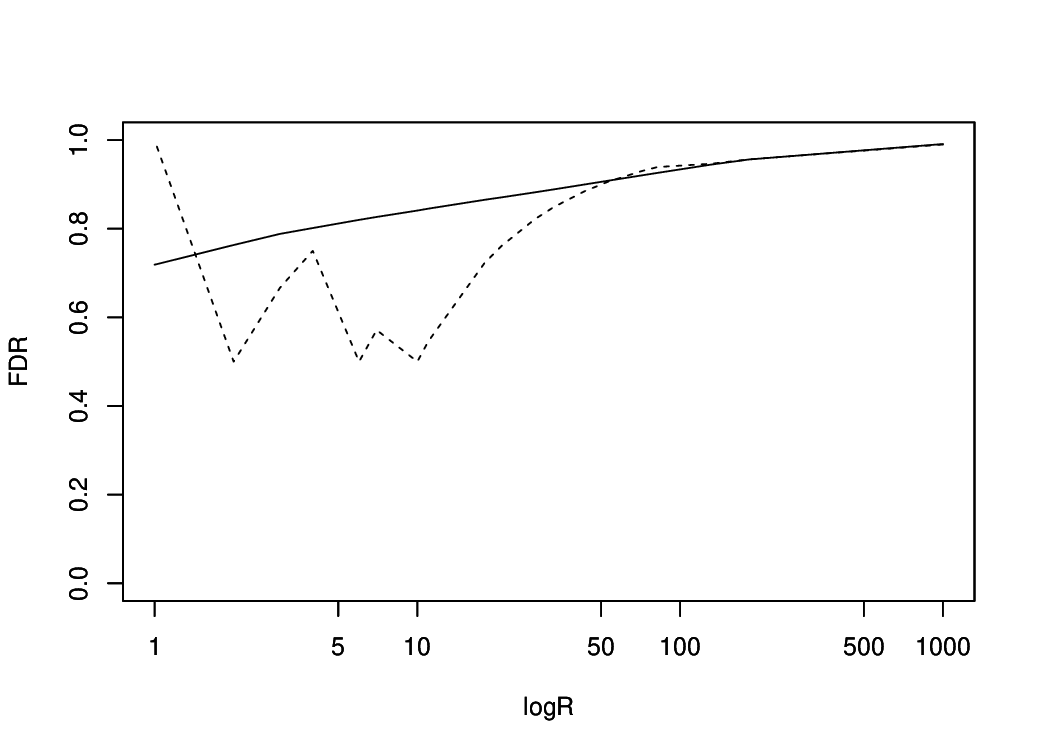}}
\caption{FDR vs log($R$). Straight and dashed lines indicate the posterior and counted FDRs, respectively.
Panels (a), (b), and (c) show the results of the model without outliers, and panels (d), (e), and (f) show the results of the model with outliers.}\vspace{0cm}
\end{center}
\label{fig:3}
\end{figure}

\section{Evaluation with the GSE14333 Dataset}
\label{sec:15}
\subsection{Data}
\label{sec:16}
The robust Bayesian modeling discussed above was applied to the gene expression data of colorectal cancer (GSE14333) available from Gene Expression Omnibus, which is provided by \cite{R19}. The GSE14333 dataset includes 54,613 genes from 290 individuals.
The gene expression data of GSE14333 were skewed to the right. Thus, we used their logarithmic values.
The state variable $\gamma_{i}$ was defined as 0 for cancer stages A and B and 1 for stages C and D.

\subsection{Results}
\label{sec:17}

Before the estimation of the posterior FDR, we examined the performance of the model by cross-validating $ppp(y)$ values on a dataset of 1,000 genes selected using the unreplaced sampling method. The presence of outliers was detected by the value of $ppp(S_g^2)$ as well. Each MCMC process was iterated 1,000 times. The first 100 iterations were removed, leaving 900 samples for analysis. The candidate distributions comprised two $t$ distributions with 3 and 15 degrees of freedom.

Figures 4 and 5 show the results of $ppp(y)$ and $ppp(S_g^2)$. Under the $t$ distribution with 15 degrees of freedom, the $ppp(y_{gi})$ was almost uniformly distributed, and the $ppp(S_g^2)$ exhibited a mode around the center, implying conservative results (see Figures 4 and 5). Thus, we selected 15 degrees of freedom of the $t$-distribution for our model.

\begin{figure}[ht]
\begin{center}
\subfloat[$ppp(S^2) \quad df=3$]{
\includegraphics[width=0.35\textwidth]{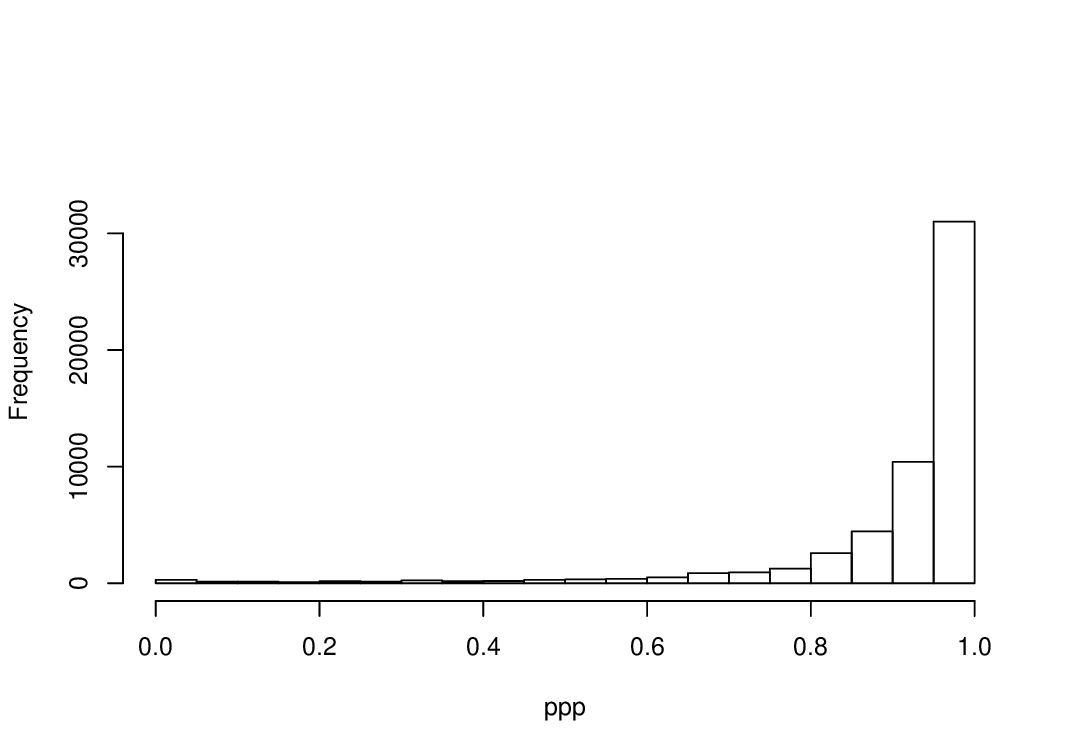}}
\subfloat[$ppp(S^2) \quad df=3$]{
\includegraphics[width=0.35\textwidth]{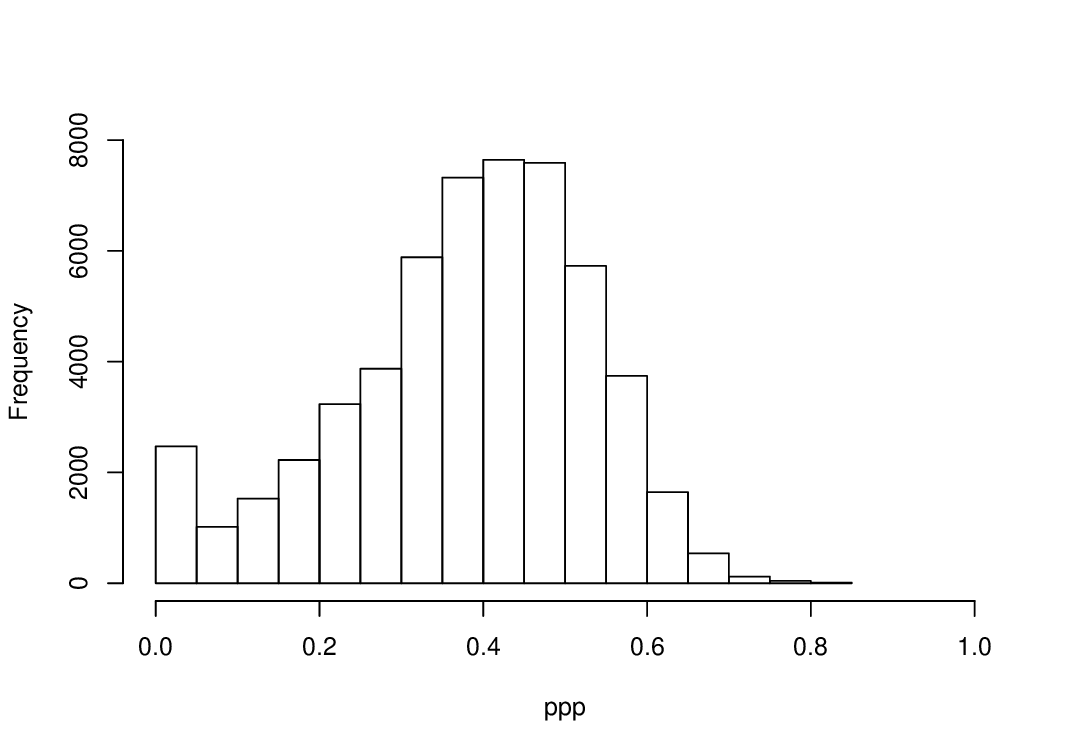}}\\
\subfloat[$ppp(y_i) \quad df=3$]{
\includegraphics[width=0.35\textwidth]{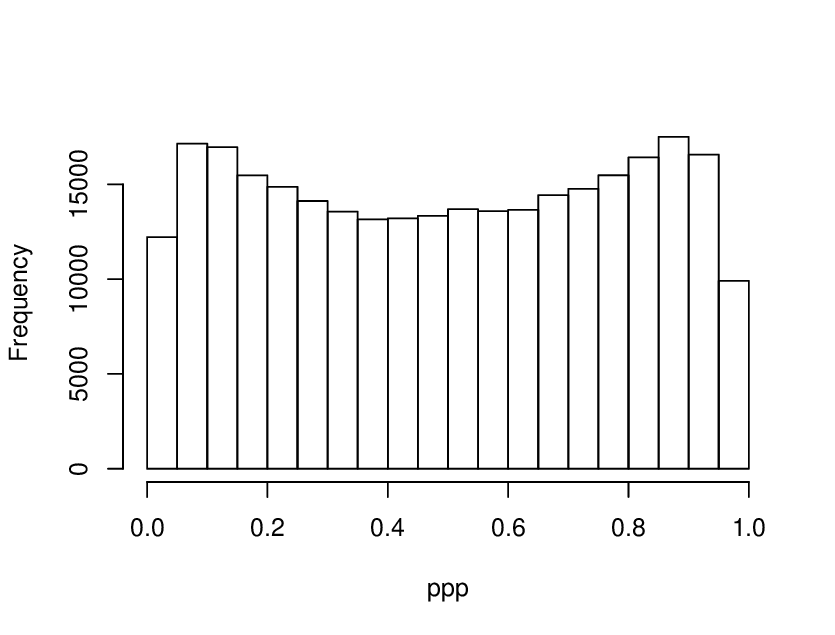}}
\subfloat[$ppp(y_i)  \quad df=15$]{
\includegraphics[width=0.35\textwidth]{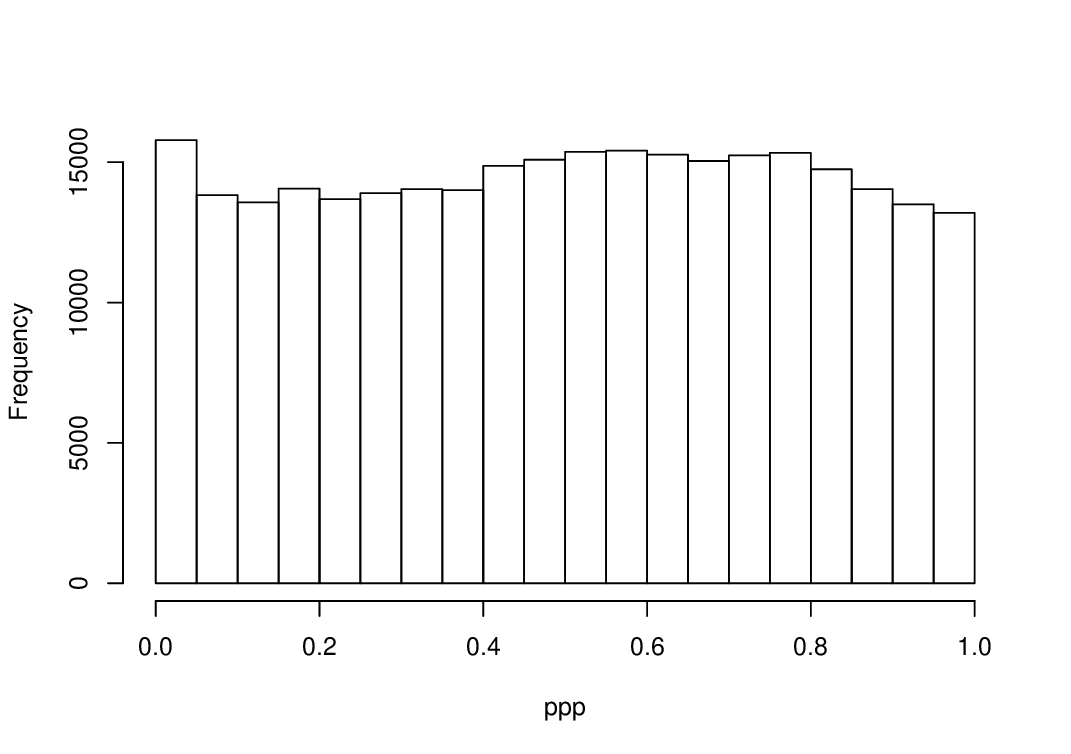}}
\caption{Posterior predictive $p$-value of $S^2$  and $y_i$ obtained by cross-validation for randomly selected 1,000 genes in GSE14333. The results of the model using $t$ distribution with (a) (c) 3 degrees of freedom and (b) (d) 15 degrees of freedom.} \label{fig4}
\end{center}
\end{figure}

\begin{figure}[ht]
\begin{center}
\includegraphics[width=0.35\textwidth]{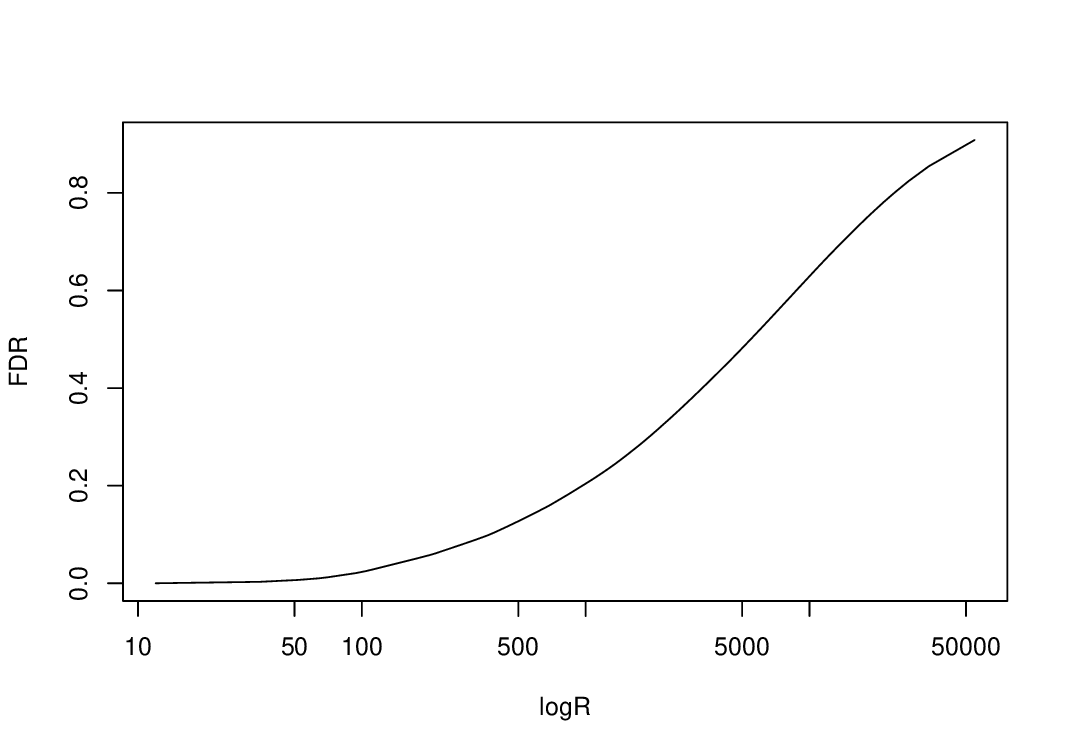}
\caption{FDR v.s. log($R$) for GSE14333} \vspace{0cm}
\end{center}
\label{fig5}
\end{figure}

 Using the model where the degrees of freedom of $t$ distribution were 15, we applied Storey's $q$-value method; the null ratio was estimated to be 0.912, which corresponded to the adaptive critical value of 0.0132. 

Figure 5 shows the result of the posterior FDR. From the result, for $p_{cut}$ = 0.5 (the posterior FDR $=$ 0.323), we found 2,189 discovery genes. When we set the posterior FDR = 0.20, we found 947 discovery genes.

\section{Discussion}
\label{sec:D}

Here, we provide a robust Bayesian analysis of large-scale datasets. 
We proposed an adaptive criterion that provides the level of the posterior FDR using the estimated null ratio and observed that it provides more reasonable results than a simple Bayesian FDR.
Moreover, we applied robust Bayesian modeling using Student $t$-modeling and diagnostics to detect outliers in differential gene expression analysis on a large scale based on the posterior predictive distribution.
Using the adaptive Bayesian FDR, we achieved effective control of the FDR level. Moreover, the results show that the heavy-tailed modeling using Student-$t$ distribution performed robustly against outliers.

However, the simulation results show that the dataset with outliers takes a larger estimated null ratio, which induces more conservative FDR. The posterior FDR also shows a more conservative FDR for the dataset with outliers. Thus, when the dataset includes outliers, we should note that our estimated posterior FDR becomes more conservative, 
\\

\clearpage

\noindent
\bf{\textsf{Disclosure statement}}

\rm{No potential conflict of interest was reported by the authors.\\

\noindent
\bf{\textsf{Grants}}

\rm{This work was partly supported by MEXT Promotion of Distinctive Joint Research Center Program JPMXP0723833165 and Osaka Metropolitan University Strategic Research Promotion Project (Development of International Research Hubs).\\

\noindent
\bf{\textsf{Supplemental data}}

\rm{Supplemental data for this article can be accessed at doi: 10.4121/adc4d0a1-52f3-4f73-850a-f5a5724bed1c.

\end{document}